\documentclass[runningheads,a4paper]{llncs}

\usepackage{amssymb}
\usepackage{subfig}
\usepackage{graphicx}
\usepackage{epstopdf}
\usepackage{multirow}
\usepackage{amsmath}
\usepackage{caption}
\usepackage{array}
\usepackage{url}

\usepackage[belowskip=-15pt,aboveskip=0pt]{caption}
\let\llncssubparagraph\subparagraph
\let\subparagraph\paragraph
\usepackage[compact]{titlesec}
\let\subparagraph\llncssubparagraph
\newcolumntype{C}[1]{>{\centering\let\newline\\\arraybackslash\hspace{0pt}}m{#1}}

\urldef{\mailsa}\path|{alfred.hofmann, ursula.barth, ingrid.haas, frank.holzwarth,|
\urldef{\mailsb}\path|anna.kramer, leonie.kunz, christine.reiss, nicole.sator,|
\urldef{\mailsc}\path|erika.siebert-cole, peter.strasser, lncs}@springer.com|
\newcommand{\keywords}[1]{\par\addvspace\baselineskip
\noindent\keywordname\enspace\ignorespaces#1}

\begin{document}
\parskip 0pt
\mainmatter  

\title{Formal Availability Analysis using Theorem Proving\thanks{ The final publication is available at http://link.springer.com}}


%
%
\author{Waqar Ahmed %
\and Osman Hasan }
%

\institute{School of Electrical Engineering and Computer Science\\
National University of Sciences and Technology (NUST),
Islamabad, Pakistan \\
\email{ \{waqar.ahmad,osman.hasan\}@seecs.nust.edu.pk  }
}

%
%

\maketitle

\begin{abstract}
Availability analysis is used to assess the possible failures and their restoration process for a given system. This analysis involves the calculation of instantaneous and steady-state availabilities of the individual system components and the usage of this information along with the commonly used availability modeling techniques, such as Availability Block Diagrams (ABD) and Fault Trees (FTs) to determine the system-level availability. Traditionally, availability analyses are conducted using paper-and-pencil methods and simulation tools but they cannot ascertain absolute correctness due to their inaccuracy limitations. As a complementary approach, we propose to use the higher-order-logic theorem prover HOL4 to conduct the availability analysis of safety-critical systems. For this purpose, we present a higher-order-logic formalization of instantaneous and steady-state availability, ABD configurations and generic unavailability FT gates. For illustration purposes, these formalizations are utilized to conduct formal availability analysis of a satellite solar array, which is used as the main source of power for the Dong Fang Hong-3 (DFH-3) satellite.

\end{abstract}

\section{Introduction}
Availability analysis is used to identify and assess the causes and frequencies of system failures. The outcomes of availability analysis play a vital role in ensuring failure-free operation of the given system. Due to the rapid increase in the usage of technological systems in safety and mission-critical domains, such as transportation and healthcare, the demand of their availability and thus availability analysis is also growing dramatically.

 The first step, in the availability analysis, is the evaluation of basic metrics of reliability and maintainability, such as mean-time to failure (MTTF) \cite{Trivedi_02}, mean-time between failure (MTBF) \cite{Trivedi_02} and mean-time to repair (MTTR) \cite{Trivedi_02}, at the individual \textit{component level} of the given system. These metrics are then used to calculate the availability of each component of the system by using the reliability and the maintainability distributions, such as \textit{Exponential} or \textit{Weibull}, with failure and repair rates, $\lambda = \frac{1}{MTTF}$ and $\mu = \frac{1}{MTTR}$. The next step is the selection of an appropriate availability modeling technique, such as Availability Block Diagrams (ABD) \cite{stapelberg2009handbook} and unavailability Fault Trees (FT) \cite{stapelberg2009handbook}. These techniques are the extension of traditionally used reliability modeling techniques, such as Reliability Block Diagram (RBD) \cite{Trivedi_02} and Fault Tree (FT) \cite{Trivedi_02},  for availability analysis purposes. Besides these two techniques, Markov chains \cite{blake1989multistage} have also been used for availability assessment. In practice, it provides much more detailed analysis compared to ABD and UFT. However, the major problem with the Markov chain based availability analysis is its exponential growth in the state-space as the system complexity increases \cite{blake1989multistage}. For instance, consider the large Multistage Interconnection Networks (MINs) \cite{blake1989multistage} that are mainly used in the supercomputers and multi-process systems to realize communication among thousands of processors. To conduct the Markov chain based availability analysis of a 8 x 8 MIN consisting of 16 switching elements, we need to consider $2^{16}$ possible states \cite{blake1989multistage}. Although, we can somewhat reduce the number of states by taking appropriate assumptions but it can compromise the accuracy of the availability results \cite{blake1989multistage}. On the other hand, ABD and UFT are intuitive and transparent methods that can be used to describe the availability of large and complex  systems, like MINs \cite{bistouni2014analyzing}.
 The  ABD and UFT based modeling techniques also allow us to estimate the availability of the given system at the \textit{system level} and play a particularly useful role at the design stages of the system to scrutinize the design alternatives without building the actual system. Once an appropriate availability model is obtained then the next step is to perform the \textit{system level} availability analysis of the model using an appropriate analysis technique.

Traditionally,  simulation tools, such as ReliaSoft \cite{Reliasoft_14} and ASENT \cite{ASENT_14}, are used to analyze the availability models. However, these techniques cannot be termed as accurate due to their inherent incompleteness and the involvement of pseudo-random numbers and numerical methods. Given the safety and financial-critical nature of many technological systems these days, a slight unavailability of such a system, at a particular instant, may lead to disastrous situations, including the loss of human lives or heavy financial setbacks. For instance, it is reported that the Amazon Web Service (AWS) suffered an unavailability for 12 hours, in April 21, 2011, causing hundreds of high-profile Web sites to go offline \cite{bailis2014network}, which resulted in a loss of 66,240 US\$ per minute downtime of its services.

 Model checking techniques have been used to overcome the above-mentioned limitations for conducting the reliability analysis (e.g.,\cite{robidoux_10,bozzano2009compass}), which is in turn used to assess the failure free operation of a system in a given interval and is thus quite closely related to availability analysis. Stochastic Petri Nets (SPN) have also been utilized to formalize RBD and FT, which are then used to analyze the availability \cite{signoret2013make}. However, a major disadvantage of using these approaches is their inability to analyze large size systems. Moreover, the computation of probabilities in these methods \cite{robidoux_10,bozzano2009compass} involves numerical methods, which compromises the accuracy of the results. Leveraging upon the high expressiveness of higher-order logic and a recent
formalization of probability theory \cite{mhamdi_11}, the higher-order-logic theorem prover HOL4 has been recently used for the formalization of Reliability Block Diagrams (RBD) \cite{WAhmad_CICM14,WAhmed_Wimob15} and Fault trees (FT) \cite{CICM_15_WAhmed}. These efforts clearly indicate the effectiveness of using a higher-order-logic (HOL) theorem prover for conducting reliability and failure analysis and, in the current paper, we develop the reasoning support for availability analysis by extending the HOL4 formalizations of RBD and FT. It is important to note that our proposed approach of using HOL theorem proving for availability analysis is primarily based on deductive reasoning. The availability properties are verified by using sound reasoning process and it is supported by the fact that every new theorem is derived from already verified theorems \cite{gordon_93}. Therefore, the analysis is much more rigorous and accurate compared to computer algebra systems (CAS), such as Mathematica \cite{mathematica}, which simplify the given closed form expressions and returns the results in the form of symbolic expressions. This fact can be illustrate with this example that the simplification of the expression $\frac{(x^{2}-1)}{(x-1)}$ by CAS yields $\mathit{(x + 1)}$ without explicitly mentioning $\mathit{(x \neq 1)}$ \cite{harrison1994extending}. On the other hand, HOL theorem prover cannot verify the same expression without this premise.

The main contribution of the paper is to formalize the ABD, unavailability FT gates and steady-state availability to develop a formal library of availability theory foundations. This library can then be used to model and analyze both component and system level availability properties of any system within the sound core of a theorem prover. The main challenge faced in this formalization, compared to our earlier formalizations related to reliability theory,  was to introduce the notion of an availability event that is associated with each system component. Each one of these availability events consists of a sequence of multiple random variables that are functioning over time. In order to illustrate the effectiveness of our proposed formalization, we present a formal availability analysis of a satellite solar array \cite{wu2011reliabilityRBD,wu2011reliability} that has been used as a main power source for the Dong Fang Hong-3 (DFH-3) satellite. In addition, we also provide some automated reasoning support for the availability analysis. This automation allows us to quantitatively compute the availability and unavailability of the DFH-3 satellite solar array from the given values of the failure and repair rates.

\section{Probability and Reliability in HOL}
\label{sec:prelim}
Mathematically, a measure space is defined as a triple ($\Omega,\Sigma, \mu$), where
$\Omega$ is a set, called the sample space, $\Sigma$ represents a $\sigma$-algebra of subsets of
$\Omega$, where the subsets are usually referred to as measurable sets, and $\mu$ is a measure with domain
$\Sigma$. A probability space is a measure space ($\Omega,\Sigma, Pr$), such that the measure,
referred to as the probability and denoted by $Pr$, of the sample space is 1. In the HOL formalization of probability theory \cite{mhamdi_11}, given a probability space $p$, the functions \texttt{space}, \texttt{subsets}  and \texttt{prob} return the corresponding $\Omega$, $\Sigma$ and $Pr$, respectively. This formalization also includes the formal verification of some of the most widely used probability axioms, which play a pivotal role in formal reasoning about reliability properties. A random variable is a measurable function between a probability space and a measurable space. The measurable functions belong to a special class of functions, which preserves the property that the inverse image of each measurable set  is also measurable. A measurable space refers to a pair ($S,\mathcal{A}$), where $S$ denotes a set and $\mathcal{A}$ represents a nonempty collection of sub-sets of $S$. Now, if $S$ is a set with finite elements, then the corresponding random variable is termed as a discrete random variable otherwise it is called a continuous one.

Now, reliability $R(t)$ is defined as the probability of a system or component performing its desired task over certain interval of time and expressed mathematically in terms of random variable as $R(t) = Pr (X > t)$. This concept can be formalized in HOL4 as follows:

\begin{flushleft}
	\label{CDF_def}
	\vspace{1pt} \small{\texttt{$\vdash$ $\forall$  p X t. Reliability p X t = distribution p X \{y | Normal t < y\}
		}}
	\end{flushleft}
	
	\noindent where the variables \texttt{$p:(\alpha \rightarrow bool) \# ((\alpha \rightarrow bool) \rightarrow bool) \# ((\alpha \rightarrow bool) \rightarrow real)$}, $X: (\alpha \rightarrow extreal)$ and $t:real$ represent a probability space, a random variable and a  $real$ number respectively. The function \texttt{Normal} takes  a $real$ number as its inputs and converts it to its corresponding value in the $extended-real$ data-type, i.e, it is the $real$ data-type with the inclusion of  positive and negative infinity. The function \texttt{distribution} takes three parameters:  a probability space $p$, a random variable $X$ and a set of $extended-real$ numbers and outputs  the probability of a random variable $X$ that acquires all the values of the given set in probability space $p$.
	
\section{Instantaneous and Steady-state Availabilities}
The instantaneous or point availability $A_{inst}(t)$ of a system or component can be defined as the probability that the given system or component is properly functioning at a given time instant $t$. If there are no repairs required after the fault has occurred then the availability \textit{A(t)} is simply equal to the reliability \textit{R(t)} of the system. However, if the system or component requires repair, then the availability can be considered as the function of two random variables, i.e., $X_{i} = T_{i} + D_{i}$, where $T_i$ is the working time
in the $i^{th}$ period and $D_i$ is the repair time in the  $i^{th}$ period. If the time when a system starts working in the $k^{th}$ period is $S_k = \sum_{i=1}^{k-1} X_i$ then the considered system is said to be available at time $t$ when there exists a period such that $S_k \le t < S_k + T_k$. Now, the corresponding availability event constituted by these random variables can be formalized in HOL4 as follows:

\begin{flushleft}
	\small{\texttt{\bf{Definition 1: }}}
	\label{avail_event_def}
	\small{\vspace{1pt} \texttt{$\vdash$ $\forall$ p X t. avail\_event p L n t =\\
			\{x |
				SIGMA ($\lambda$a. FST (EL a L) x + SND (EL a L) x) (count n) $\le$ t $\wedge$\\ \
				t <
				SIGMA ($\lambda$a. FST (EL a L) x + SND (EL a L) x) (count n) +\\ \ \ \ \ \ \ \
				FST (EL n L) x\} $\cap$ p\_space p
		}}
	\end{flushleft}
	
\noindent The above definition takes a probability space $p$, a list of random variable pairs $L$, representing the working and repair time random variables, a number $n$ and a time variable $t$ and returns the corresponding availability event. The function \texttt{SIGMA} takes an arbitrary function $f$ and a set $s$ and returns the sum of all the values obtained by applying the function $f$ on each element of the given set. The HOL4 function \texttt{count} takes a number $n$ and returns a set containing all the natural numbers less than the given number $n$. Similarly, the function \texttt{EL} takes an index variable and a list and retrieves the list element located at the given index number. The HOL4 functions \texttt{FST} and \texttt{SND} are primarily used to access the first and second elements in a pair. Definition 1 models the  corresponding event of the $i^{th}$ working interval only. To cover all the working  intervals, we take the union of these availability events, corresponding to the pairs of random variable in list $L$, in HOL4 as follows:

\begin{flushleft}
	\small{\texttt{\bf{Definition 2: }}}
	\label{avail_event_def}
	\small{\vspace{1pt} \texttt{$\vdash$ $\forall$ p L t.
			union\_avail\_events p L t =\\
			BIGUNION
			(IMAGE ($\lambda$a. avail\_event p L a t) (count (LENGTH L)))
		}}
	\end{flushleft}

 An interesting property of the availability event is that its probability, also known as instantaneous availability, is always greater or equal to the corresponding reliability, i.e., $R_{T_{1}}(t) \le A_{inst}(t)$, where $T_{1}$ is the first time-to-work random variable. This property can be formally verified, based on Definitions 1 and 2, in HOL4 as follows:

 \begin{flushleft}
 	\small{\texttt{\bf{Theorem 1: }}} \label{avail_ge_rel}
 	\vspace{1pt} \small{\texttt{$\vdash$ $\forall$ p t L. prob\_space p $\wedge$
 			(0 $\le$ t) $\wedge$ $\neg$NULL L $\wedge$\\
 			($\forall$n. avail\_event p L n t $\in$ events p)  $\wedge$ \\
 			($\forall$a b.
 			(a $\neq$ b) $\Rightarrow$\\
 			\ DISJOINT (avail\_event p L a t)
 			(avail\_event p L b t))
 			 $\Rightarrow$\\
 			\ \ (Reliability p (FST (HD L)) t $\le$  prob p (union\_avail\_events p L t))
 		}}
 	\end{flushleft}

\noindent The first two assumptions ensure that $p$ is a valid probability space and time index $t$ must be positive. The next two assumptions make sure that the given list of random variables must not be empty and the availability events are in the events space $p$.  The last assumption ensures that the availability events are disjoint. The conclusion models the property that the instantaneous availability is always greater or equal to reliability. The function \texttt{Reliability} takes a probability space $p$, a random variable that is associated with the system or component and a time variable $t$ and returns the reliability of the system or component \cite{WAhmad_CICM14}.

 Consider that the failure and repair random variables are exhibiting exponential distributions with failure and repair rates $\lambda$ and $\mu$, respectively, then the instantaneous availability at the component level can be expressed mathematically as follows \cite{Trivedi_02}:
{\small \begin{equation}\label{eq:inst_avail}
\begin{split}
   & A_{inst}(t) = \frac{\mu}{\mu + \lambda} + \frac{\lambda}{\mu + \lambda} e^{-(\lambda + \mu)t}
\end{split}
\end{equation}
}

\noindent where the failure and repair rates are the mean-time-to-failure (MTTF) and mean-time-to-repair (MTTR), i.e. $\lambda$ = $\frac{1}{MTTF}$ and $\mu$ = $\frac{1}{MTTR}$, which are basic metrics for reliability and maintainability, respectively.

 Now, we can formalize the instantaneous availability, given in Equation \ref{eq:inst_avail}, as follows:
\begin{flushleft}
\small{\texttt{\bf{Definition 3: }}}
\label{inst_avail_def}
\vspace{1pt} \small{\texttt{$\vdash$ $\forall$ p L m.
		inst\_avail\_exp p L m =\\
		$\forall$t.
		prob p (union\_avail\_events p L (\&t)) =\\ \vspace{2mm}
		$\dfrac{\texttt{SND m}}{\texttt{(SND m + FST m)}}$  +
		$\dfrac{\texttt{FST m}}{\texttt{(SND m + FST m)}}$ * exp (-(SND m + FST m) * \&t)
}}
\end{flushleft}

\noindent where the variables \texttt{FST m} and \texttt{SND m} represent failure and repair rates, respectively.

The steady-state availability of any component, which reflects the long-term availability after the system becomes stable, can be evaluated by taking the limit as $t$ approaches infinity in Equation (\ref{eq:inst_avail}).
{\small \begin{equation}\label{eq:steady_avail}
  A_{steady} = \lim_{t \rightarrow \infty}A_{inst}(t) = \frac{\mu}{\mu + \lambda}
\end{equation}}

The above equation can be formally verified in HOL4 as follows:
\begin{flushleft}
\small{\texttt{\bf{Theorem 2: }}} \label{inst_avail_THM}
\vspace{1pt} \small{\texttt{$\vdash$ $\forall$ p L m.
		prob\_space p $\wedge$ (0 < FST m $\wedge$ 0 < SND m) $\wedge$\newline
		($\forall$t.
		($\forall$a b.
		a $\neq$ b $\Rightarrow$\\
		DISJOINT (avail\_event p L a t)
		(avail\_event p L b t)) $\wedge$\\
		($\forall$n. avail\_event p L n t $\in$ events p)) $\wedge$
		 inst\_avail\_exp p L m $\Rightarrow$\\ \vspace{2mm}
	\ \	(lim
		($\lambda$t. prob p (union\_avail\_events p L (\&t))) =
		$\dfrac{\texttt{SND m}}{\texttt{(SND m + FST m))}}$
 }}
\end{flushleft}

\noindent The assumptions of the above theorem are quite similar to those used in Theorem 1. The proof of Theorem 2 is primarily based on the fact that the negative exponential function tends to zero as its exponent tends to infinity.

\section{Availability Block Diagrams}
\label{sec:ABD}
 Availability Block Diagram (ABD) are graphical structures that represent the system components and their interconnections in the form of blocks and connector lines, respectively. The system is termed as available, if at least one path of properly available components from the input to output exists.

\begin{figure}[t!]
\centering
\subfloat[]{\includegraphics[width=2in]{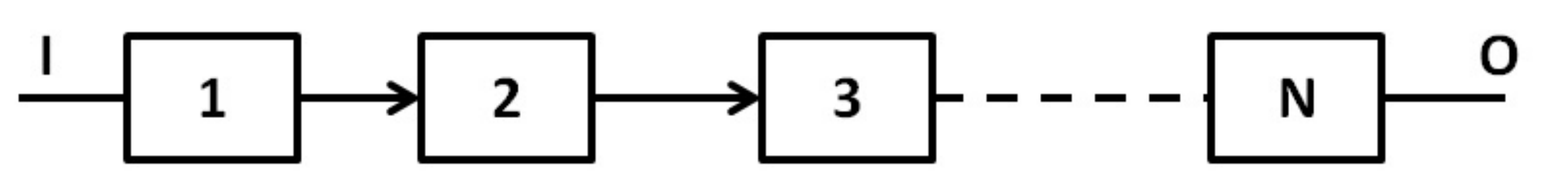}}
\subfloat[]{\includegraphics[width=1.1in]{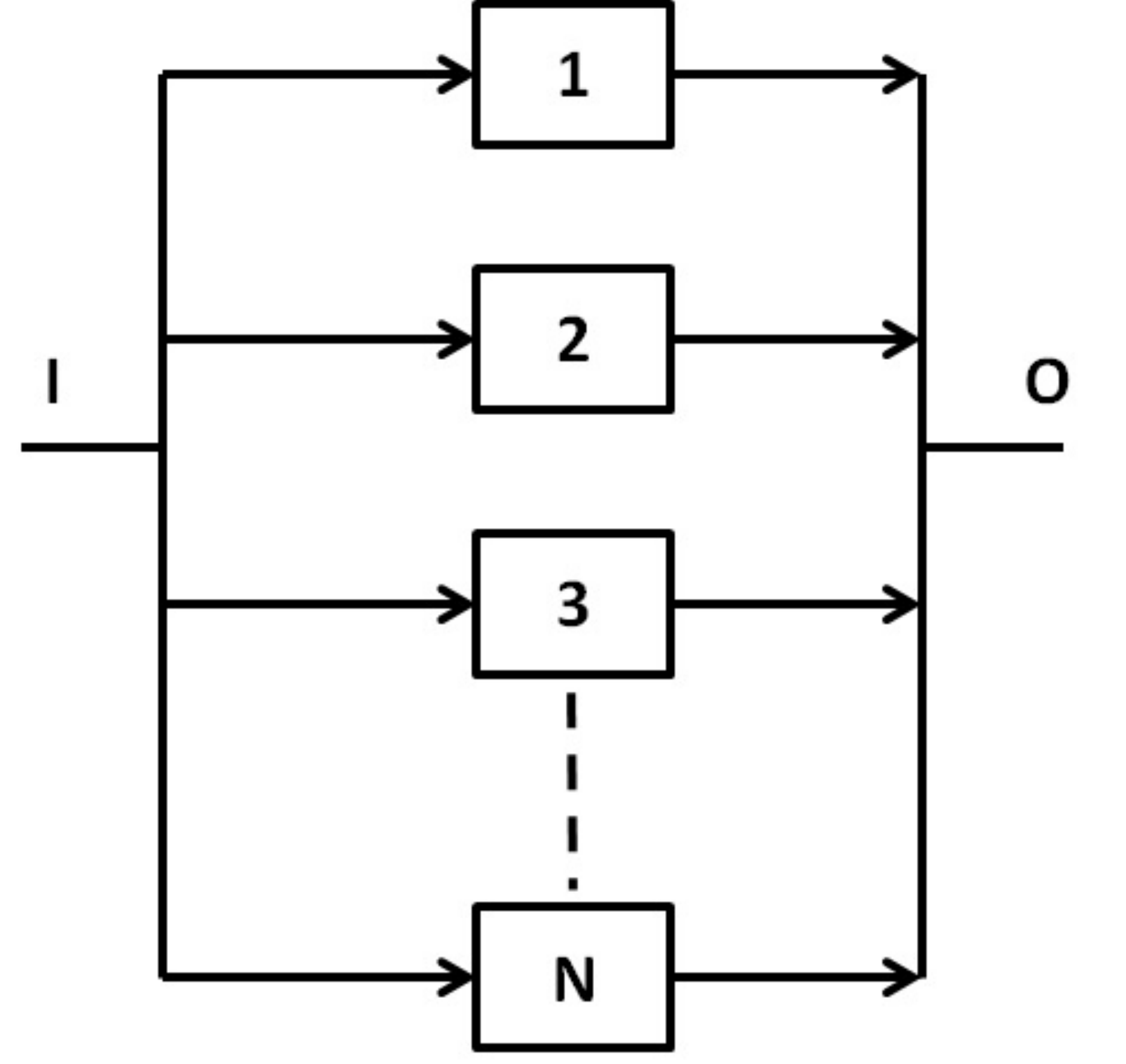}}\vspace{8pt}
\newline
\subfloat[]{\includegraphics[width=2in]{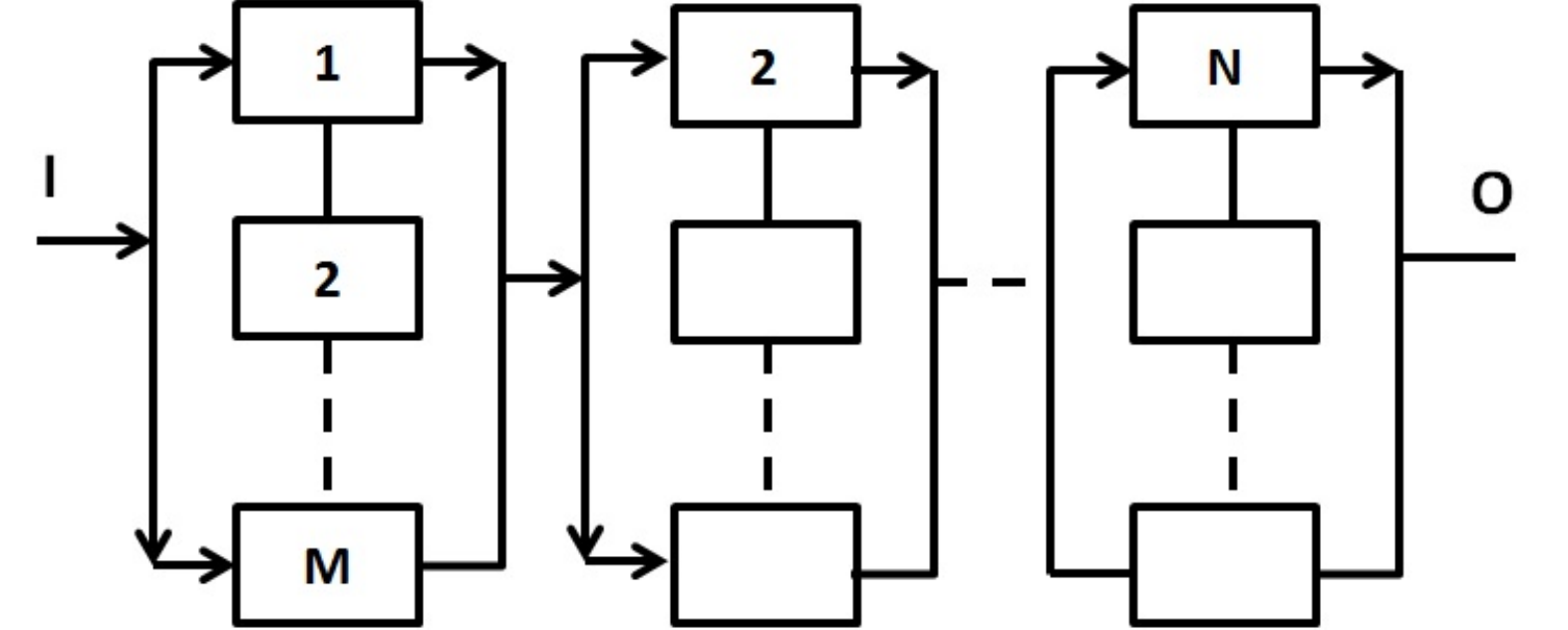}}\hspace{10pt}
  \subfloat[]{\includegraphics[width=1.7in]{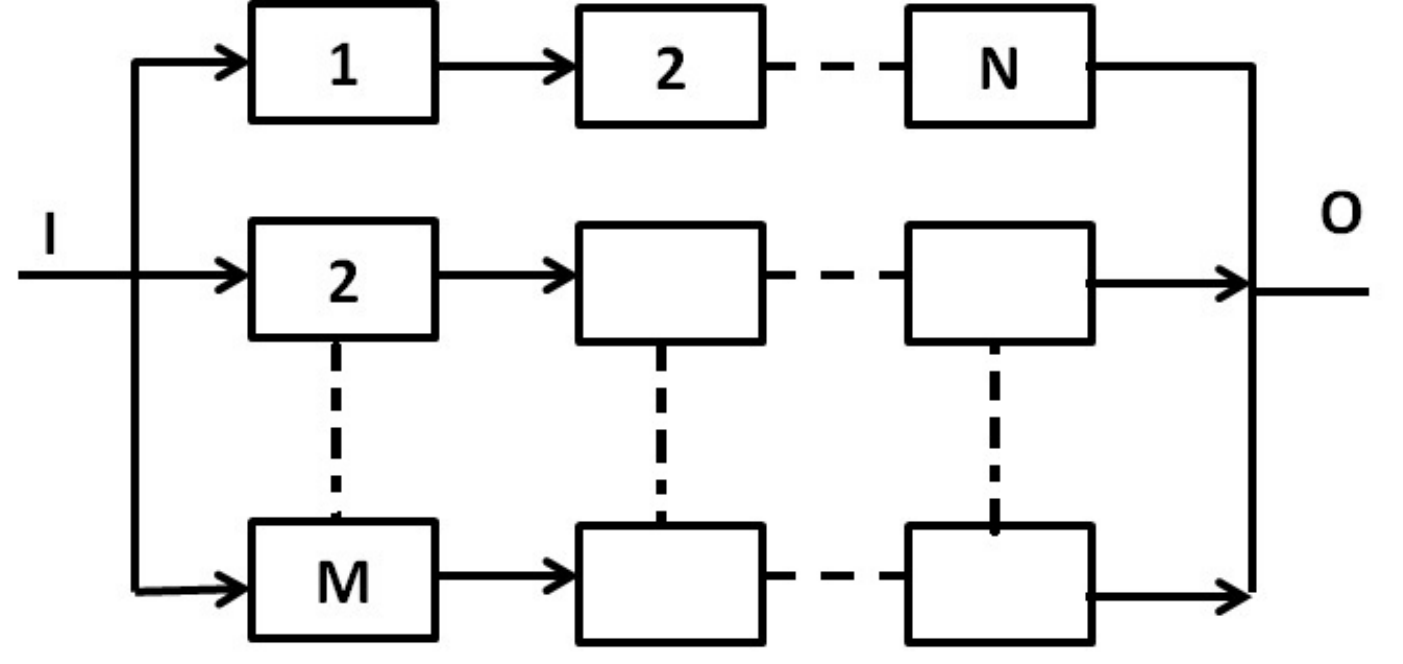}}
\caption{ABDs (a) Series (b) Parallel (c) Series-Parallel (d) Parallel-Series}
\label{fig:RBDs}
\end {figure}

The availability of a system with components connected in series is considered to be available at time instant $t$ only if all of its components are available at time $t$, as depicted in Figure \ref{fig:RBDs}(a). If $A_{inst_{i}}(t)$ is a mutually independent event that represents the instantaneous availability of the $i^{th}$ component of a serially connected system with $N$ components at time instant $t$, then the steady-state availability of the complete system can be expressed as \cite{ebeling2004introduction}:
{\small
\begin{equation}\label{eq3:series_ABD}
\begin{split}
       \lim_{t \rightarrow \infty}Pr (\bigcap_{i=1}^{N}A_{inst_{i}}(t))
       = \prod_{i=1}^{N}(\frac{\mu_i}{\mu_i + \lambda_i})
      \end{split}
 \end{equation}}

The series ABD configuration can be formalized as:
	\begin{flushleft}
		\small{\texttt{\bf{Definition 4: }}}
		\label{series_def}
		\vspace{1pt} \small{\texttt{$\vdash$ ($\forall$ p.
				series\_struct p []  = p\_space p) $\wedge$\\
				($\forall$ p h t. series\_struct p (h::t) = h $\cap$ series\_struct p t)
			}}
		\end{flushleft}

\noindent The above function takes a list of events corresponding to the availability of individual components of the given system and the probability space $p$ and returns the intersection of all of the elements in a given list and the whole probability space, if the given list is empty. Based on this definition, Equation (\ref{eq3:series_ABD}) can be formally verified as follows:

\begin{flushleft}
	\small{\texttt{\bf{Theorem 3: }} \label{series_connected_ABD_system_THM}
		\vspace{1pt} \texttt{$\vdash$ $\forall$ p L M.
			(A1): prob\_space p $\wedge$ (A2): (0 $\le$ t) $\wedge$\newline (A3): ($\forall$z. MEM z M $\Rightarrow$ 0 < FST z $\wedge$  0 < SND z) $\wedge$ \newline
			(A4): (LENGTH L = LENGTH M) $\wedge$\newline
			(A5): ($\forall$t'.
			$\neg$NULL (union\_avail\_event\_list p L (\&t')) $\wedge$ \newline
			(A6): ($\forall$z. MEM z (union\_avail\_event\_list p L (\&t')) $\Rightarrow$ z $\in$ events p) $\wedge$\newline
			(A7): mutual\_indep p (union\_avail\_event\_list p L (\&t'))) $\wedge$\newline
			(A8): inst\_avail\_exp\_list p L M $\Rightarrow$\\
			\ (lim
			($\lambda$t.
			prob p (series\_struct p (union\_avail\_event\_list p L (\&t)))) =\\
			\ \ list\_prod (steady\_state\_avail\_list M))
		}}
	\end{flushleft}

\noindent where the function \texttt{union\_avail\_event\_list} can be obtained by mapping the function \texttt{union\_avail\_event} on every element of the given random variable list. The function \texttt{list\_prod} returns the product of given real number list. The first two assumptions (A1-A2) ensure that $p$ is a valid probability space and the time $t$ must be positive. The assumptions (A3-A4) guarantee that the failure and repair rates are positive and the length of failure-repair random variable and the corresponding rate lists are equal. The next two assumptions (A5-A6) make sure that the length of availability event list, representing the availability of individual components, must not be empty and each availability event in a \texttt{avail\_event\_list} is in events space $p$. The last two assumptions (A7-A8) provide the mutual independence among all the availability events and the  instantaneous availability of each component. The conclusion of the theorem represents Equation (\ref{eq3:series_ABD}) as the function \texttt{steady\_state\_avail\_list} takes a list of pairs, representing the failure and repair rates, and returns a list of steady-state availabilities, corresponding to each component of the given system.

Similarly, the availability of a system with parallel connected components, depicted in Figure \ref{fig:RBDs}(b), mainly depends on the component with the maximum availability. In other words, the system will continue functioning as long as at least one of its components remains functional.
 Mathematically \cite{ebeling2004introduction}:
{\small
\begin{equation}\label{eq4:parallelRBD}
   \lim_{t \rightarrow \infty}Pr (\bigcup_{i=1}^{N}A_{inst_{i}}(t) ) = 1 - \prod_{i=1}^{N}(1 - \frac{\mu_i}{\mu_i + \lambda_i})
      \end{equation}
      }
Now, the availability of a system with a parallel structure is defined as:

	\begin{flushleft}
		\small{\texttt{\bf{Definition 5: }}} \label{parallel_def}
		\vspace{1pt} \small{\texttt{$\vdash$ (parallel\_struct [] = \{\}) $\wedge$\\
				($\forall$ h t. parallel\_struct (h::t) =  h $\cup$ parallel\_struct t)
			}}\end{flushleft}
			
			The function \texttt{parallel\_struct} accepts a list of reliability events and returns the parallel structure reliability event by recursively performing the union operation on the given list of reliability events  or an empty set if the given list is empty.
We can now verify Equation (\ref{eq4:parallelRBD})  as follows:
\begin{flushleft}
\small{\texttt{\bf{Theorem 4: }} \label{parallel_connected_ABD_system_THM}
\vspace{1pt} \texttt{$\vdash$ $\forall$p L M.\\
	\ (lim
	($\lambda$t.
	prob p (parallel\_struct p (union\_avail\_event\_list p L (\&t)))) =\\
	\ \  1 -  list\_prod (one\_minus\_list (steady\_state\_avail\_list M))
}}
\end{flushleft}
The above theorem is verified under the same assumptions as Theorem 3. The conclusion of the theorem represents Equation (\ref{eq4:parallelRBD}) where, the function \texttt{one\_minus\_} \texttt{list} accepts a list of $real$ numbers $[x1, x2, \cdots, xn]$ and returns the list of $real$ numbers such that each element of this list is 1 minus the corresponding element of the given list, i.e., $[1-x1, 1-x2 \cdots, 1-xn]$.
The proof of Theorem 4 is based on Theorem 3 along with the fact that given a list of $n$ mutually independent events, the complement of these $n$ events are also mutually independent.

If in each serial stage the components are connected in parallel, as shown in Figure \ref{fig:RBDs}(c), then the configuration is termed as a series-parallel structure. If $A_{inst_{ij}}(t)$ is the event corresponding to the instantaneous availability of the $j^{th}$ component connected in an  $i^{th}$ subsystem at time instant $t$, then the steady-state availability of the complete system can be expressed as follows \cite{ebeling2004introduction}:
{\small
\begin{equation}\label{eq:series-parallelABD}
   \lim_{t \rightarrow \infty}Pr (\bigcap_{i=1}^{N} \bigcup_{j=1}^{M} A_{inst_{ij}}(t))=  \prod_{i=1}^{N}(1 - \prod_{j=1}^{M} (1- \frac{\mu_{ij}}{\mu_{ij} + \lambda_{ij}}))
\end{equation}}

By extending the ABD formalization approach, presented in Theorems 3 and 4, we formally verify the generic availability expression for series-parallel ABD configuration,  given in Equation (\ref{eq:series-parallelABD}), in HOL4 as follows:

\begin{flushleft}
\small{\texttt{\bf{Theorem 5: }} \label{series_parallel_connected_system_THM}
\vspace{1pt} \texttt{$\vdash$ $\forall$
            p L M.
            prob\_space p  $\wedge$ (LENGTH L = LENGTH M) $\wedge$\\
            ($\forall$z. MEM z (FLAT M) $\Rightarrow$ 0 < FST z $\wedge$ 0 < SND z) $\wedge$
            \\
            ($\forall$n. n < LENGTH L $\Rightarrow$ (LENGTH (EL n L) = LENGTH (EL n M))) $\wedge$\\
            ($\forall$t'.
            ($\forall$z. MEM z (list\_union\_avail\_event\_list p L (\&t')) $\Rightarrow$ $\neg$NULL z) $\wedge$\\
            ($\forall$z'.
            MEM z' (FLAT (list\_union\_avail\_event\_list p L (\&t'))) $\Rightarrow$\\
            \ z' $\in$ events p) $\wedge$\\
            mutual\_indep p (FLAT (list\_union\_avail\_event\_list p L (\&t')))) $\wedge$\\
            two\_dim\_inst\_avail\_exp p L M $\Rightarrow$\\
            (lim
            ($\lambda$t.
            prob p \\ \ \
            (series\_parallel\_struct p
            (list\_union\_avail\_event\_list p L (\&t)))) =\\
            \ list\_prod
            (one\_minus\_list (MAP ($\lambda$a. compl\_steady\_state\_avail a) M)))
}}
\end{flushleft}

\noindent where the function \texttt{list\_union\_avail\_event\_list} is obtained by mapping the function \texttt{union\_avail\_event\_list} on each element of the given random variable list.

 The function \texttt{series\_parallel\_struct} models the series-parallel ABD by first mapping the function \texttt{parallel\_struct} on each element of the given event list and then applying the function \texttt{series\_struct} to this obtained list. Similarly, the function  \texttt{compl\_steady\_state\_avail} returns a list of one minus steady-state availabilities.

The functions \texttt{list\_prod} and \texttt{one\_minus\_list}
are used to model the product and complement of steady-state availabilities, respectively. The assumptions are similar to the ones used in Theorems 3 and 4 with the extension that the given lists are two-dimensional lists.
The HOL4 function \texttt{FLAT} is used to convert a two dimensional list into a single list. The  conclusion models the right-hand-side of Equation (\ref{eq:series-parallelABD}). The proof of the above theorem uses Theorems 3 and 4 and also requires a lemma that given the list of mutually independent reliability events,  an event corresponding  to the series-parallel structure and a reliability event are also independent in probability.

  If the components in these reserved \emph{subsystems} are connected serially then the structure is called a parallel-series structure, as depicted in Figure \ref{fig:RBDs}(d). If $A_{ij}(t)$ is the event corresponding to the availability of the $j^{th}$ component connected in a  $i^{th}$ subsystem at time $t$, then the steady-state availability becomes:
  {\small
 \begin{equation}\label{eq:parallel-seriesABD}
 \lim_{t \rightarrow \infty}Pr (\bigcup_{i=1}^{M} \bigcap_{j=1}^{N} A_{ij}(t))=1- \prod_{i=1}^{M}(1 - \prod_{j=1}^{N} \frac{\mu_{ij}}{\mu_{ij} + \lambda_{ij}} )
\end{equation}
}The above equation is also verified as a HOL4 theorem in our development and more details about it can be found in \cite{waqar_ABD_ITP_15}.

\section{Unavailability Fault Trees}
\label{sec:UFT}
 Unavailability FT is a graphical technique consisting of internal nodes, which are represented by gates like OR, AND and XOR, and the external nodes, that model the unavailability events, which are associated with the occurrence of faults in components of the given system. The generic nature of these gates allows us to construct an efficient and accurate unavailability fault tree (FT) model for any given system. This FT can in turn be used to investigate the potential causes of a fault occurrence, which makes the system unavailable, and the calculation of minimal number of unavailability events, known as minimal cut-set (MCS), that contribute towards the occurrence of a $top$ $event$, i.e., a critical event, which can cause the whole system unavailable upon its occurrence.

 We can formalize the unavailability event of a system by taking the complement of the availability event with respect to the probability space \textit{p}.
 \begin{flushleft}
\small{\texttt{\bf{Definition 6: }}}
\label{avail_event_def}
\small{ \texttt{$\vdash$$\forall$ p X t.\\ union\_unavail\_events p L t =
		p\_space p DIFF union\_avail\_events p L t
}}
\end{flushleft}

The instantaneous unavailability of the system can be expressed as follows:
{\small\begin{equation}\label{eq:inst_unavail}
  \overline{A_{inst}}(t)  = \frac{\lambda}{\mu + \lambda} - \frac{\lambda}{\mu + \lambda} e^{-(\lambda + \mu)t}
\end{equation}}
The HOL4 formalization of the above equation is as follows:
\begin{flushleft}
\small{\texttt{\bf{Definition 7: }}}
\label{inst_avail_def}
\vspace{1pt} \small{\texttt{$\vdash$ $\forall$  p L m.
		inst\_unavail\_exp p L m =\\
		$\forall$t.
		prob p (union\_unavail\_events p L (\&t)) =\\ \vspace{2mm}
		$\dfrac{\texttt{FST m}}{\texttt{(SND m + FST m)}}$  -
		$\dfrac{\texttt{FST m}}{\texttt{(SND m + FST m)}}$ * exp (-(SND m + FST m) * \&t)
}}
\end{flushleft}

 If the occurrence of the unavailability event at the output is caused by the occurrence of all the input unavailability events then this kind of behavior can be modeled by using the AND unavailability FT gate, as shown in Table \ref{table:FT_gate_def}.

{\small
\begin{equation}\label{eq:and_unavail_gate}
      Pr (\bigcap_{i=2}^{N}\overline{A_{inst_{i}}}(t))
      = \prod_{i=2}^{N}\frac{\lambda_{i}}{\lambda_{i} + \mu_{i}}
 \end{equation}
}

 The above equation can be formalized in HOL4 as follows:

\begin{flushleft}
	\small{\texttt{\bf{Theorem 6: }} \label{series_connected_ABD_system_THM}
		\vspace{1pt} \texttt{$\vdash$ $\forall$ p L M.
			prob\_space p $\wedge$\\ ($\forall$z. MEM z M $\Rightarrow$ 0 < FST z $\wedge$  0 < SND z) $\wedge$
			(LENGTH L = LENGTH M) $\wedge$\newline
			($\forall$t'.
			$\neg$NULL (union\_unavail\_event\_list p L (\&t')) $\wedge$ \\
			($\forall$z. MEM z (union\_unavail\_event\_list p L (\&t')) $\Rightarrow$ z $\in$ events p) $\wedge$\\
			mutual\_indep p (union\_unavail\_event\_list p L (\&t'))) $\wedge$\\
			inst\_unavail\_exp\_list p L M $\Rightarrow$\\
			\ (lim
			($\lambda$t.\\ \ \ \ \
			prob p (AND\_unavail\_FT\_gate p (union\_avail\_event\_list p L (\&t)))) =\\
			\ \ list\_prod (steady\_state\_unavail\_list M))
		}}
	\end{flushleft}
	
\begin{table}[t!]
	\centering
	\caption{HOL Formalization of Fault Tree Gates}
	\scalebox{1}{
		\begin{tabular}{|l |l|}
			\hline
			\shortstack{\small{Unavail.} \\ \small{FT Gates}} & HOL Formalization  \\
			\hline
			\hline
			\parbox[s]{0.5em}{
				\includegraphics[width=0.5in]{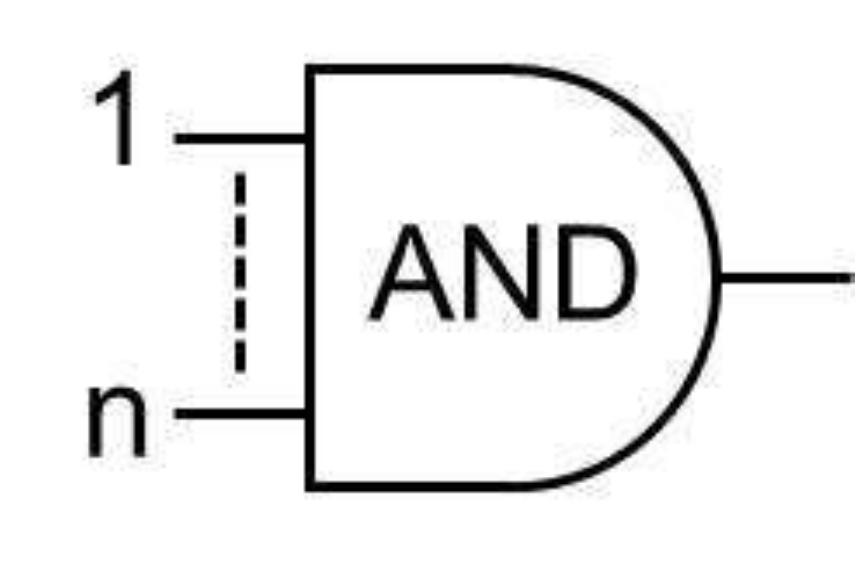}}
			& \small{\texttt{$\vdash$ $\forall$ p L t.
					AND\_unavail\_FT\_gate p L t =}}\\
					& \ \ \texttt{inter\_list p (union\_unavail\_event\_list p L t)}
			\\
			\parbox[c]{0.5em}{
				\includegraphics[width=0.5in]{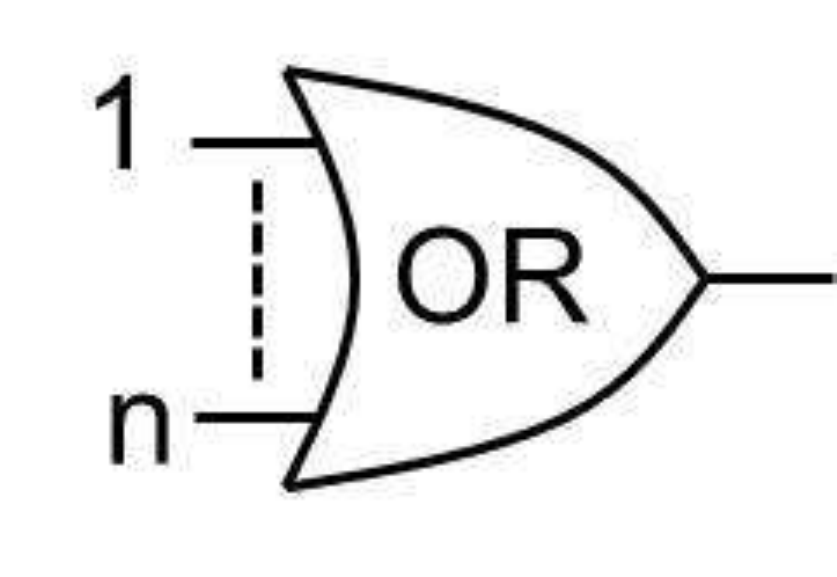}} &  \small{\texttt{$\vdash$ $\forall$ p L t.
					OR\_unavail\_FT\_gate p L t =}}\\
					& \ \ \ \small{\texttt{union\_list (union\_unavail\_event\_list p L t)
				}} \\
				\parbox[l]{0.5em}{
					\includegraphics[width=0.63in, trim={1.8cm 0 0 0},clip]{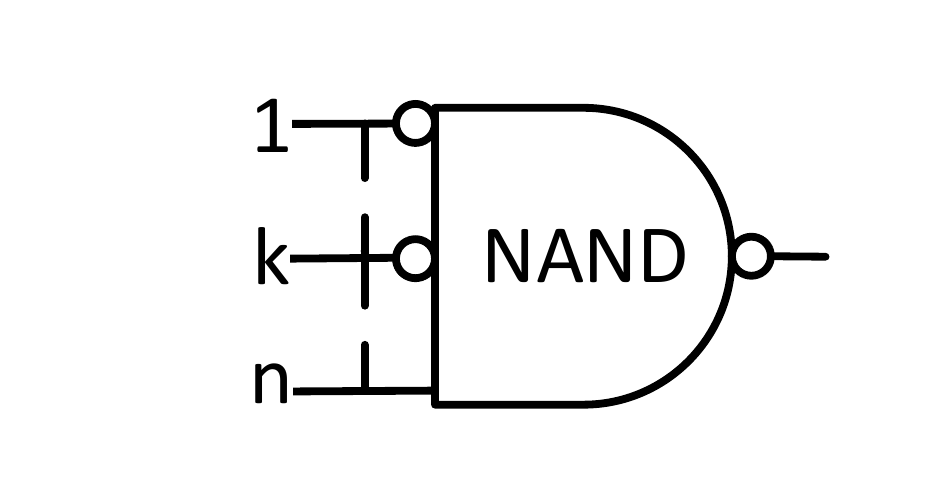}} & \small{\texttt{$\vdash$ $\forall$p L1 L2 t.
							NAND\_unavail\_FT\_gate p L1 L2 t =}}\\
						&	\small{\texttt{inter\_list p (compl\_list p (union\_unavail\_event\_list p L1 t)) $\cap$}}\\
							&	\small{\texttt{inter\_list p (union\_unavail\_event\_list p L2 t)
					}} \\
					\parbox[l]{0.5em}{
						\includegraphics[width=0.5in]{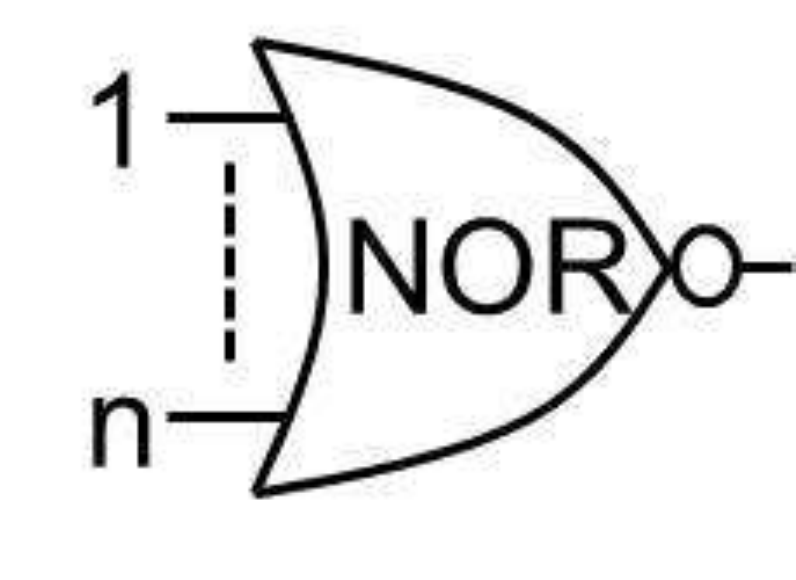}}& \small{\texttt{$\vdash$ $\forall$ p L t.
							NOR\_unavail\_FT\_gate p L t =}}\\
						&	\small{\texttt{p\_space p DIFF union\_list (union\_unavail\_event\_list p L t)}
					} \\
					\parbox[t]{0.5em}{
						\includegraphics[width=0.5in]{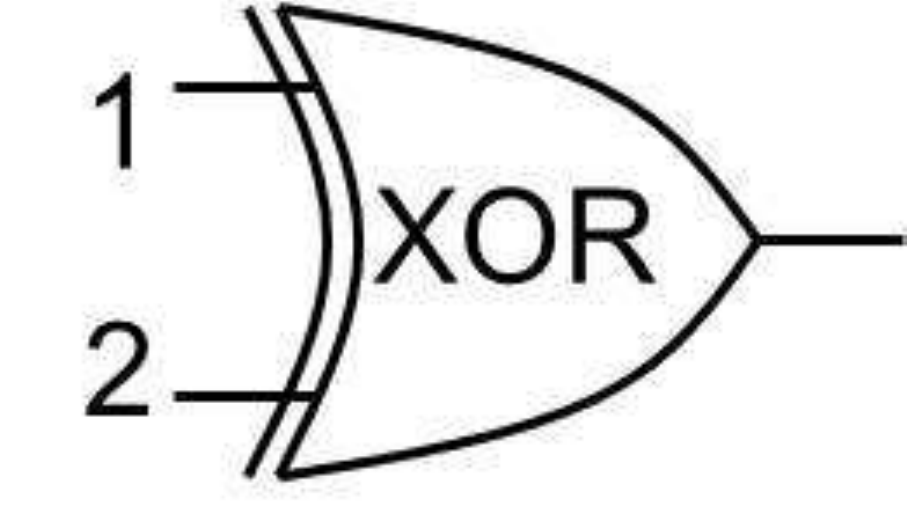}} & \small{\texttt{$\vdash$ $\forall$ p A B. XOR\_FT\_unavail\_gate p A B =}}\\
					 &\small{\texttt{((p\_space p DIFF A $\cap$ B) $\cup$ (A $\cap$ p\_space p DIFF B))
							}} \\
							\parbox[l]{0.5em}{
								\includegraphics[width=0.5in]{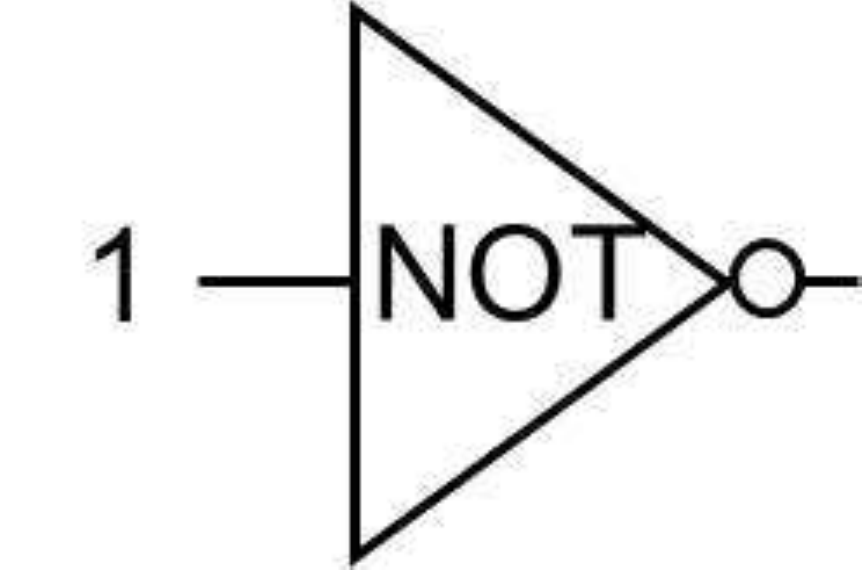}}& \small{\texttt{$\vdash$ $\forall$ p A. NOT\_unavail\_FT\_gate p A =  (p\_space p DIFF A)
								}} \\
								\hline
							\end{tabular}}\label{table:FT_gate_def}
						\end{table}
						
 \noindent The assumptions of the above theorem are similar to the ones used in Theorem 2 and the conclusion of Theorem 5 represents Equation (\ref{eq:and_unavail_gate}).

 In the OR unavailability FT gate, the occurrence of the output unavailability event depends upon the occurrence of any one of its input unavailability event. The  function \texttt{OR\_unavail\_FT\_gate}, given in Table 1, models this behavior as it returns the union of the input unavailability list $L$ by using the recursive function \texttt{union\_list}. The NOR unavailability FT gate, modeled by using the function \texttt{NOR\_unavail\_FT\_gate}, given in Table 1, can be viewed as the complement of the OR unavailability FT gate and its output unavailability event occurs if none of the input unavailability event occurs.

 Similarly, the NAND unavailability FT gate, represented by the function \texttt{NAND\_unavail\_FT\_gate} in Table 1, models the behavior of the occurrence of an output unavailability event when at least one of the unavailability events at its input does not occur. This type of gate is used in unavailability FTs when the non-occurrence of the unavailability event in conjunction with the other unavailability events causes the top unavailability event to occur. This behavior can be expressed as the intersection of complementary and normal events, where the complementary events model the non-occurring unavailability events and the normal events model the occurring unavailability events. The output unavailability event occurs in the 2-input XOR unavailability FT gate if only one, and not both, of its input unavailability events occur. The HOL4 representation of the behaviour of the \texttt{XOR\_unavail\_FT\_gate} is also presented in Table \ref{table:FT_gate_def}. The function \texttt{NOT\_unavail\_FT\_gate} accepts an unavailability event $A$ and probability space $p$ and returns the complement to the probability space $p$ of the given input unavailability event $A$. The verification of the corresponding unavailability expressions, of the above-mentioned unavailability FT gates, is presented in Table \ref{table:unavail_FT_thms}. These expressions are verified under the same assumptions as the ones used for Theorem 6 and the proofs are mainly based on some fundamental mutual independence properties of the given unavailability events along with some axioms of probability theory.
{\small
 \begin{table}[!htb]
\centering
\caption{Unavailability Fault Tree Gates}
\scalebox{0.93}{
\begin{tabular}{|l|l|}
\hline
Unavailability FT Gates & Conclusions of the formally verified Theorems \\
\hline
\hline
$\!\begin{aligned}[t]
 \lim_{t \rightarrow \infty}\overline{A}_{OR}(t) = \lim_{t \rightarrow \infty}Pr (\bigcup_{i=1}^{N}A_{inst_{i}}(t) )\\
    = 1 - \prod_{i=2}^{N}(1 - \frac{\lambda_{i}}{\lambda_{i} + \mu_{i}})&
    \end{aligned}$  &
 $\!\begin{aligned}[t] & \small{\texttt{ lim ($\lambda$t. prob p}} \\ & \ \small{\texttt{(OR\_unavail\_FT\_gate p L \&t)  =}} \\ &
 \small{\texttt{1 - list\_prod (one\_minus\_list} }\\ & \ \small{\texttt{(steady\_state\_unavail\_list M)))
}}  \end{aligned}$ \\
\hline

$\!\begin{aligned}[t]
  \lim_{t \rightarrow \infty}\overline A_{NOR}(t)
     = 1 -  \lim_{t \rightarrow \infty}\overline{A}_{OR}(t)&  \\
    = \prod_{i=2}^{N}(1 - \frac{\lambda_{i}}{\lambda_{i} + \mu_{i}})&
    \end{aligned}$  &
 $\!\begin{aligned}[t] & \small{\texttt{ (lim ($\lambda$t. prob p}} \\ & \ \small{\texttt{ (NOR\_unavail\_FT\_gate p L \&t))  =}} \\ &
  \small{\texttt{list\_prod (one\_minus\_list} }\\ & \ \small{\texttt{(steady\_state\_unavail\_list M}}   \end{aligned}$ \\
\hline

$\!\begin{aligned}[t]
 \lim_{t \rightarrow \infty}\overline{A}_{NAND}(t)  = & \\ \lim_{t \rightarrow \infty}Pr (\bigcap_{i=2}^{k}A_{i}(t) \cap \bigcap_{j=k}^{N}\overline A_{i}(t)) = & \\ \prod_{i=2}^{k}(1 - \frac{\mu_{i}}{\mu_{i} + \lambda_{i}}) *   \prod_{j=k}^{N}\frac{\lambda_{i}}{\mu_{i} + \lambda_{i}} & \end{aligned}$  &  $\!\begin{aligned}[t]& \small{\texttt{(lim ($\lambda$t. prob p}} \\ & \ \small{\texttt{(NAND\_unavail\_FT\_gate p L1 L2 t)  = }}\\ &
 \small{\texttt{list\_prod (steady\_state\_avail M1) *}} \\ & \small{\texttt{ list\_prod (steady\_state\_unavail\_list M2}}  \end{aligned}$ \\
\hline

$\!\begin{aligned}[t]
\lim_{t \rightarrow \infty}\overline A_{XOR}(t)&=  \\ \lim_{t \rightarrow \infty}Pr(\bar{A}(t)B(t) \cup A(t)\bar{B}(t)) = & \\ (1- \frac{\lambda_{1}}{\lambda_{1} + \mu_{1}})*  \frac{\lambda_{2}}{\lambda_{2} + \mu_{2}} +    \frac{\lambda_{1}}{\lambda_{1} + \mu_{1}} * \\ (1- \frac{\lambda_{2}}{\lambda_{2} + \mu_{2}}) &\end{aligned}$  &
$\!\begin{aligned}[t]& \small{\texttt{(lim ($\lambda$t.  prob p }} \\ & \ \small{\texttt{(XOR\_unavail\_FT\_gate p A B \&t))  =}}\\ & \small{\texttt{(1 - (steady\_state\_unavail M1))}}* \\ & \ \small{\texttt{(steady\_state\_unavail M2) + }}\\ & \small{\texttt{(steady\_state\_unavail M1)}} * \\ &  \ \small{\texttt{(1 - (steady\_state\_unavail M2))}} \end{aligned}$ \\
\hline

$\!\begin{aligned}[t]
\lim_{t \rightarrow \infty}\overline A_{NOT}(t)= Pr(A(t)) =(1 - \frac{\lambda}{\lambda + \mu})&\end{aligned}$  & $\!\begin{aligned}[t] & \small{\texttt{lim ($\lambda$t. prob p (NOT\_FT\_gate p A \&t)}} = \\ & \small{\texttt{ FST m / (FST m + SND m)}} \end{aligned}$  \\
\hline
\end{tabular}}\label{table:unavail_FT_thms}
\end{table}
}

 The principle of inclusion exclusion (PIE) forms an integral part of the reasoning involved in verifying the unavailability of a FT. In FT based unavailability analysis, firstly all the basic  unavailability events are identified that can cause the occurrence of the system top unavailability event. These unavailability events are then combined to model the overall fault behavior of the given system by using the fault gates. These combinations of basic  unavailability events, called cut sets, are then reduced to  minimal cut sets (MCS) by using set-theory rules, such as idempotent, associative and commutative. The PIE is then used to evaluate the overall failure probability of the given system.

If $\overline A_{i}$ represent the $i^{th}$ basic unavailability event or a combination of  unavailability events then the overall unavailability of the given system can be expressed in terms of the probabilistic inclusion-exclusion principle as follows:
{\small
 \begin{equation}\label{PIE}
\mathbb{P} (\bigcup_{i=1}^n \overline A_i)  = \sum_{J \neq \{\}, J\subseteq\{1,2,\ldots,n\}}(-1)^{|J|-1} \mathbb{P} (\bigcap_{j\in J} \overline A_j)
 \end{equation}
 }

\noindent The above equation has been formalized in HOL4 as follows \cite{CICM_15_WAhmed}:
 \begin{flushleft}
\small{\texttt{\bf{Theorem 7: }}} \label{PIE_THM}
\vspace{1pt} \small{\texttt{$\vdash$ $\forall$ p L t. prob\_space p $\wedge$ \\ ($\forall$ x. MEM x (union\_avail\_event\_list p L t) $\Rightarrow$ x $\in$ events p) $\Rightarrow$ \\
\qquad     (prob p (union\_list (union\_avail\_event\_list p L t)) = \\ \qquad
      sum\_set \{y | y  $ \subseteq $ set (union\_avail\_event\_list p L t) $ \wedge $ y $ \neq $ \{\}\}\\ \qquad \qquad
        ($ \lambda $t. -1 pow (CARD y - 1) * prob p (BIGINTER y)))
}}
\end{flushleft}

\noindent The function \texttt{sum\_set}  recursively sums the return value of the function $f$, which is applied on each element of the given set $s$. In the above theorem, the set $s$ is represented by the term $\{x|C(x)\}$ that contains all the values of $x$, which satisfy condition $C$. Whereas, the $\lambda$ abstraction function \texttt{($ \lambda $t. -1 pow (CARD t - 1) * prob p (BIGINTER t))} models $(-1)^{|J|-1} \mathbb{P} (\bigcap_{j\in J} \overline A_j)$, such that the functions \texttt{CARD} and \texttt{BIGINTER} return the number of elements and the intersection of all the elements of the given set, respectively.

The proof script \cite{waqar_ABD_ITP_15} of the above-mentioned formalizations of ABD and unavailability FT gates and the PIE principle is composed of more than 9000 lines of HOL script and took about 350 man-hours. The main outcome of this formalization is that the definitions and theorems of ABDs and FT gates can be used to capture the behavior of wide variety of real-world systems and analyze their corresponding availability in higher-order logic.

\section{Application: Satellite Solar Arrays}

As an illustrative application to demonstrate the effectiveness of our availability theory related formalization, we consider a solar array that has been used in the DFH-3 Satellite, which was launched by the People's Republic of China on May 12, 1997 \cite{wu2011reliabilityRBD,wu2011reliability}.  Solar arrays are one of the most vital components of the satellites because the mission success heavily depends upon the continuous reliable source of power. The satellite's solar array is a mechanical system, which mainly consists of various mechanisms, including: deployable, synchronization, locking and orientation.

The solar array can be modeled by using series-parallel ABD configurations, shown in Figure \ref{RO_ABD}, and based on the availability of its individual components, such as electric detonator (ED), the cutting knife (CK), the starting spring (SS), hing bearing (HB) and hing of locking mechanism (HL), the overall availability of the solar array can be evaluated \cite{wu2011reliabilityRBD}. The HOL4 formalization of the solar array ABD (Figure \ref{RO_ABD}) is as follows:

\begin{figure}[!ht]
  \centering
  \includegraphics[width=\textwidth]{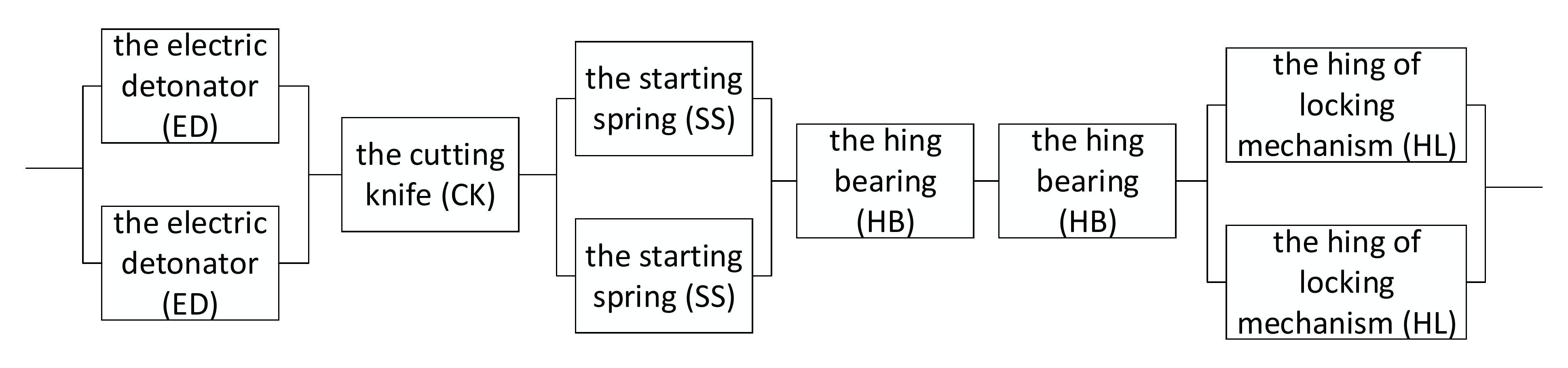}\\
  \caption{Solar Array ABD}\label{RO_ABD}
\end{figure}

\begin{flushleft}
\small{\texttt{\bf{Definition 8: }}}
\label{avail_event_def}
\small{\vspace{1pt} \texttt{$\vdash$ $\forall$p X\_ED X\_CK X\_SS X\_HB X\_HL  t. \\
RO\_ABD p X\_ED X\_CK X\_SS X\_HB X\_HL t = \\
  series\_parallel\_struct p\\
  \ \ \  (list\_union\_avail\_event\_list\\ \qquad \qquad ([[X\_ED;X\_ED];[X\_CK];[X\_SS;X\_SS];[X\_HB];[X\_HB];[X\_HL;X\_HL]]) t)
}}
\end{flushleft}
\noindent  We verified the following theorem for the availability of the satellite solar array:
\begin{flushleft}
\small{\texttt{\bf{Theorem 8 : }} \label{series_parallel_connected_system_THM}
\vspace{1pt} \texttt{$\vdash$ $\forall$p X\_ED X\_CK X\_SS X\_HB X\_HL. \\
           (lim
              ($\lambda$t.
                 prob p ( RO\_ABD p X\_ED X\_CK X\_SS X\_HB X\_HL  \&t)) = \\
            (1 - (1 - steady\_state\_avail ED) pow 2) * steady\_state\_avail CK *\\
            (1 - (1 - steady\_state\_avail SS) pow 2) *\\ ((steady\_state\_avail HB) pow 2) * (1 - (1 - steady\_state\_avail HL) pow 2)
}}

\end{flushleft}

\noindent We have omitted the assumptions of this theorem here due to space limitations and the complete formalization is available at \cite{waqar_ABD_ITP_15}. The proof of the above theorem is primarily based on Theorem 5 and is very straightforward.

An unavailability FT can be constructed by considering the faults in the solar array mechanical components, which are the fundamental causes of satellite' solar array mechanisms failure. The unavailability FT for the solar array of the DFH-3 Satellite that was launched by the People's Republic of China on May 12, 1997 \cite{wu2011reliability} is depicted in Figure \ref{RO_FT} and we formally analyze this FT in this paper. The proposed FT formalization (functions \texttt{OR\_unavail\_FT\_gate} and \texttt{AND\_unavail\_FT\_gate}, given in Table \ref{table:FT_gate_def}) is used to model the MCS of the unavailability of the solar array as follows:

\begin{figure}[!ht]
  \centering
  \includegraphics[height=8cm,width=9cm]{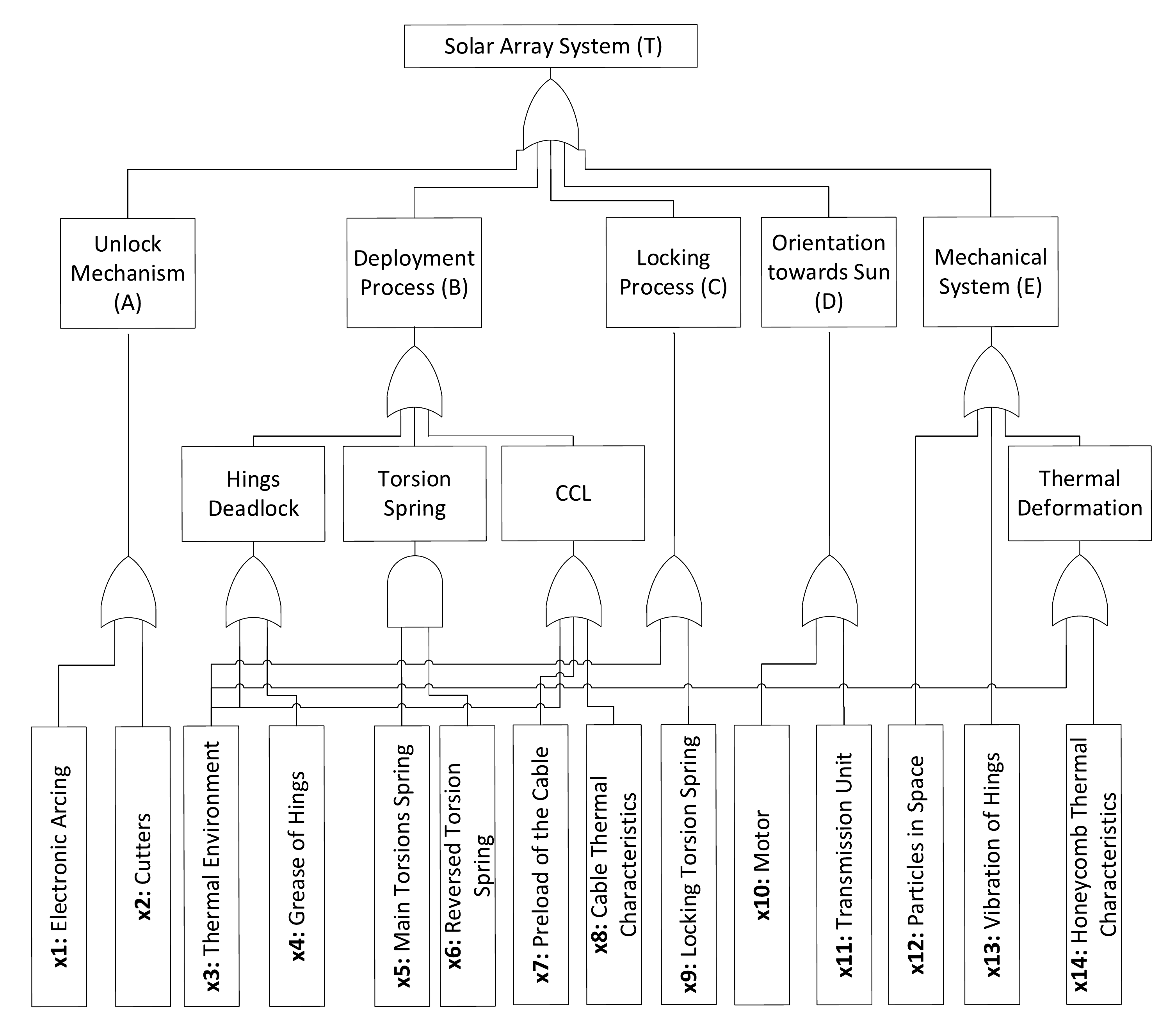}\\
  \caption{Solar Array Unavailability FT}\label{RO_FT}
\end{figure}

\begin{flushleft}\small{
		\texttt{\bf{Definition 9: }}
		\label{solar_FT_CS_def}
		\vspace{1pt} \texttt{$\vdash$ $\forall$ p x1 x2 x3 x4 x5 x6 x7 x8 x9 x10 x11 x12 x13 x14 t. \\
			Solar\_unavail\_FT p x1 x2 x3 x4 x5 x6 x7 x8 x9 x10 x11 x12 x13 x14 t = \\
			OR\_unavail\_FT\_gate\\ \qquad
			[OR\_unavail\_FT\_gate  (union\_avail\_event\_list p [x1; x2; x3; x4] t);\\ \qquad \ AND\_unavail\_FT\_gate  p (union\_avail\_event\_list p [x5; x6] t);\\
			\qquad  \     OR\_unavail\_FT\_gate \\  \qquad \quad  \ (union\_avail\_event\_list p [x7; x8; x9; x10; x11; x12; x13; x14] t)]
		}}
	\end{flushleft}

The overall unavailability of a solar array can now be verified as follows:
\vspace{3mm}
\begin{flushleft}
\small{\texttt{\bf{Theorem 9: }}} \label{solar cell_THM}
\vspace{1pt} \small{\texttt{$\vdash$  $\forall$ p x1 x2 x3 x4 x5 x6 x7 x8 x9 x10 x11 x12 x13 x14. \\
      (lim($\lambda$t.\\
      \ \  Solar\_unavail\_FT p x1 x2 x3 x4 x5 x6 x7 x8 x9 x10 x11 x12 x13 x14 \&t)) =  \\
         1 - (list\_prod (steady\_state\_unavail\_list [x5;x6]) * \\
         (1 - list\_prod (one\_minus\_list (steady\_state\_unavail\_list\\ \qquad \qquad [c1;c2;c3;c4;c6;c7;c8;c9;c10;c11;c12;c13;c14]))))
}}
\end{flushleft}

\noindent Again all quantifiers and the assumptions of the above theorem have not been included due to space limitations and the complete theorem can be found at \cite{waqar_ABD_ITP_15}. The proof of the above theorem utilizes the PIE principle (Theorem 7) and the unavailability FT gates with their corresponding mathematical expression, given in Tables \ref{table:FT_gate_def} and \ref{table:unavail_FT_thms}.

The proof script \cite{waqar_ABD_ITP_15} for Theorems 8 and 9 is composed of about 100 lines of HOL code compared to about 9000 lines of code that had to be written to formalize the foundational availability concepts. This straightforward reasoning clearly indicates the usefulness of our work. The distinguishing features of the formally verified Theorems 8 and 9, compared to the other existing availability analysis alternatives, include their generic nature, i.e., all the variables are universally quantified and thus can be specialized to obtain the availability for any given failure and repair rates, and their guaranteed correctness due to the involvement of a sound theorem prover in their verifications.
Moreover, the usage of a theorem prover in their verification ensures that all the required assumptions for the validity of the results are explicitly included in the theorems, which is quite important for designing accurate systems.

In order to facilitate the use of our formally verified results by industrial design engineers for their availability analysis, we have also developed a set of SML scripts to automate the simplification step of these theorems for any given failure and repair rate values corresponding to the DFH-3 satellite solar array components. For instance, the  \texttt{auto\_solar\_RBD\_avail} script automatically computes the availability up to 12 decimal places based on Theorem 8 as follows:
\begin{flushleft}
\vspace{1pt} \small{\texttt{$\vdash$ prob\_space p $\wedge$\\
($\forall$t'.
   ($\forall$z.
      MEM z
        (FLAT
           (list\_union\_avail\_event\_list\\ \ \
              [[X\_ED;X\_ED];[X\_CK];[X\_SS;X\_SS];[X\_HB];[X\_HB];[X\_HL;X\_HL]] (\&t'))) $\Rightarrow$\\
      z $\in$ events p) $\wedge$\\
   mutual\_indep p
     (FLAT\\
        (list\_union\_avail\_event\_list\\ \ \
           [[X\_ED;X\_ED];[X\_CK];[X\_SS;X\_SS];[X\_HB];[X\_HB];[X\_HL;X\_HL]] (\&t')))) $\wedge$\\
two\_dim\_inst\_avail\_exp p
  [[X\_ED;X\_ED];[X\_CK];[X\_SS;X\_SS];[X\_HB];[X\_HB];[X\_HL;X\_HL]]
  [[(0.1,0.3);(0.1,0.3)];[(0.2,0.5)]; [(0.3,0.4); (0.3,0.4)];
   [(0.7,0.8)]; [(0.7,0.8)]; [(0.5,0.5); (0.5,0.5)]] $\Rightarrow$ \\
\ lim
  ($\lambda$t.
     prob p ( RO\_ABD p X\_ED X\_CK X\_SS X\_HB X\_HL  \&t)) = 0.116618075802}}																																\end{flushleft}
																																
	\noindent This  \texttt{auto\_solar\_RBD\_avail} script can be used for any values of the failure and repair rates and can be easily extended to be used for the instantiation of the generic result of Theorems 9 \cite{waqar_ABD_ITP_15}. With a very little modification, these kind of automation scripts can facilitate industrial design engineers to accurate determine the availability of many other safety-critical systems.

\section{Conclusion}

The foremost requirements to conduct the formal availability analysis within a theorem prover is to formalize the ABD configurations, i.e., series, parallel, series-parallel and parallel-series, unavailability FT gates, such as AND, OR, NAND, NOR, XOR and NOT, and instantaneous and steady-state availability. This paper fulfills the above-mentioned requirement and thus provides a framework, which can be used to carry out the formal availability analysis of any system within a sound core of HOL4 theorem prover. For illustration, our formalizations are utilized to conduct the formal availability analysis of an satellite solar array and the results have been found to more rigorous than the existing availability analysis alternatives. However, this formalization is only limited to static ABD and UFT models and cannot express the time varying system states, dependent systems and non-series-parallel topologies. This limitation can be removed by extending the present formalization to dynamic ABD and dynamic UFT. This can be done by combining this formalization of ABD and UFT with the recently proposed Markov chain formalization \cite{liu2013formalization} in HOL4.

\bibliographystyle{splncs}
\bibliography{biblio}

\begin{thebibliography}{10}

\bibitem{Trivedi_02}
Trivedi, K.S.:
\newblock Probability and {S}tatistics with {R}eliability, {Q}ueuing and
  {C}omputer {S}cience {A}pplications. 2nd edn.
\newblock John Wiley and Sons Ltd. (2002)

\bibitem{stapelberg2009handbook}
Stapelberg, R.F.:
\newblock Handbook of {R}eliability, {A}vailability, {M}aintainability and
  {S}afety in {E}ngineering {D}esign.
\newblock Springer Science \& Business Media (2009)

\bibitem{blake1989multistage}
Blake, J.T., Trivedi, K.S.:
\newblock {Multistage Interconnection Network Reliability}.
\newblock Transactions on Computers \textbf{38}(11) (1989)  1600--1604

\bibitem{bistouni2014analyzing}
Bistouni, F., Jahanshahi, M.:
\newblock {Analyzing the Reliability of Shuffle-exchange Networks using
  Reliability Block Diagrams}.
\newblock Reliability Engineering \& System Safety \textbf{132} (2014)  97--106

\bibitem{Reliasoft_14}
{R}elia{S}oft:
\newblock http://www.reliasoft.com/ (2016)

\bibitem{ASENT_14}
{ASENT}:
\newblock https://www.raytheoneagle.com/asent/rbd.htm (2016)

\bibitem{bailis2014network}
Bailis, P., Kingsbury, K.:
\newblock The network is reliable.
\newblock Queue \textbf{12}(7) (2014) ~20

\bibitem{robidoux_10}
Robidoux, R., Xu, H., Xing, L., Zhou, M.:
\newblock {Automated {M}odeling of {D}ynamic {R}eliability {B}lock {D}iagrams
  {U}sing {C}olored {P}etri {N}ets}.
\newblock {IEEE Transactions on Systems, Man and Cybernetics, Part A: Systems
  and Humans} \textbf{40}(2) (2010)  337--351

\bibitem{bozzano2009compass}
Bozzano, M., Cimatti, A., Katoen, J.P., Nguyen, V.Y., Noll, T., Roveri, M.:
\newblock The {COMPASS} {A}pproach: {C}orrectness, {M}odelling and
  {P}erformability of {A}erospace {S}ystems.
\newblock In: Computer Safety, Reliability, and Security. Volume 5775 of LNCS.
\newblock Springer (2009)  173--186

\bibitem{signoret2013make}
Signoret, J.P., Dutuit, Y., Cacheux, P.J., Folleau, C., Collas, S., Thomas, P.:
\newblock {Make your Petri Nets Understandable: Reliability Block Diagrams
  Driven Petri Nets}.
\newblock Reliability Engineering \& System Safety \textbf{113} (2013)  61--75

\bibitem{mhamdi_11}
Mhamdi, T., Hasan, O., Tahar, S.:
\newblock {O}n the {F}ormalization of the {L}ebesgue {I}ntegration {T}heory in
  {HOL}.
\newblock In: Interactive {T}heorem {P}roving. Volume 6172 of {LNCS}.
\newblock Springer (2011)  387--402

\bibitem{WAhmad_CICM14}
Ahmed, W., Hasan, O., Tahar, S., Hamdi, M.S.:
\newblock Towards the {F}ormal {R}eliability {A}nalysis of {O}il and {G}as
  {P}ipelines.
\newblock In: Intelligent Computer Mathematics. Volume 8543 of LNCS.
\newblock Springer (2014)  30--44

\bibitem{WAhmed_Wimob15}
Ahmed, W., Hasan, O., Tahar, S.:
\newblock {Formal Reliability Analysis of Wireless Sensor Network Data
  Transport Protocols using HOL}.
\newblock In: Wireless and Mobile Computing, Networking and Communications,
  IEEE (2015)  217--224

\bibitem{CICM_15_WAhmed}
Ahmed, W., Hasan, O.:
\newblock {Towards Formal Fault Tree Analysis Using Theorem Proving}.
\newblock In: Conferences on Intelligent Computer Mathematics. Volume 9150 of
  LNCS.
\newblock Springer (2015)  39--54

\bibitem{gordon_93}
Gordon, M., Melham, T.:
\newblock Introduction to {H}OL: A {T}heorem {P}roving {E}nvironment for
  {H}igher-{O}rder {L}ogic.
\newblock Cambridge {P}ress (1993)

\bibitem{mathematica}
Mathematica:
\newblock www.wolfram.com (2008)

\bibitem{harrison1994extending}
Harrison, J., Th{\'e}ry, L.:
\newblock Extending the {HOL} theorem prover with a computer algebra system to
  reason about the reals.
\newblock In: Higher Order Logic Theorem Proving and Its Applications. Volume
  780 of LNCS.
\newblock Springer (1994)  174--184

\bibitem{wu2011reliabilityRBD}
Wu, H.C., Wang, C.J., Liu, P.:
\newblock {Reliability Analysis of Deployment Mechanism of Solar Arrays}.
\newblock Applied Mechanics and Materials \textbf{42} (2011)  139--142

\bibitem{wu2011reliability}
Wu, J., Yan, S., Xie, L.:
\newblock Reliability {A}nalysis {M}ethod of a {S}olar {A}rray by using {F}ault
  {T}ree {A}nalysis and {F}uzzy {R}easoning {P}etri {N}et.
\newblock Acta Astronautica \textbf{69}(11) (2011)  960--968

\bibitem{ebeling2004introduction}
Ebeling, C.E.:
\newblock An {I}ntroduction to {R}eliability and {M}aintainability
  {E}ngineering.
\newblock Tata McGraw-Hill Education (2004)

\bibitem{waqar_ABD_ITP_15}
{A}hmed, W.:
\newblock Formalization of {A}vailability {B}lock {D}iagram and
  {U}navailability {FT}.
\newblock http://save.seecs.nust.edu.pk/availability/ (2016)

\bibitem{liu2013formalization}
Liu, L.Y.:
\newblock {Formalization of Discrete-time Markov Chains in HOL}.
\newblock PhD thesis, Concordia University (2013)

\end{thebibliography}
\end{document}